\documentclass{sig-alternate}

\usepackage[ruled,lined,longend]{algorithm2e}
\usepackage[usenames]{color}
\usepackage{gastex}
\usepackage{listings}
\usepackage{url}
\usepackage{subfigure}

\newcommand{\event}{\ensuremath{e}}
\newcommand{\eventset}{\ensuremath{E}}
\newcommand{\initset}{\ensuremath{I}}

\newcommand{\tlen}{\ensuremath{len}}
\newcommand{\ttop}{\ensuremath{top}}

\newcommand{\abstractTC}{abstract event sequence}

\newcommand{\abstractTCs}{abstract event sequences}
\newcommand{\AbstractTCs}{Abstract event sequences}

\newcommand{\execTC}{executable event sequence}

\newcommand{\execTCs}{executable event sequences}
\newcommand{\ExecTCs}{Executable event sequences}

\newcommand{\tc}{\ensuremath{s}}
\newcommand{\testset}{\ensuremath{S}}
\newcommand{\pathset}{\ensuremath{\Pi}}
\newcommand{\efgedge}{\ensuremath{\delta}}
\newcommand{\edgedge}{\ensuremath{\psi}}
\newcommand{\weight}{\ensuremath{w}}

\newtheorem{definition}{Definition}

\definecolor{color-comment}{rgb}{0,0.50,0}
\definecolor{color-string}{rgb}{0.64,0.08,0.08}
\definecolor{color-background}{rgb}{0.96,0.96,0.96}

\begin{document}

\conferenceinfo{ISSTA}{'2012 Minneapolis, Minnesota USA}

\title{Grey-box GUI Testing:\\Efficient Generation of Event Sequences}

\numberofauthors{5}
\author{
\alignauthor
Stephan Arlt\\
       \affaddr{Institut f\"ur Informatik}\\
       \affaddr{Albert-Ludwigs-Universit\"at Freiburg}\\
       \email{arlt@informatik.uni-freiburg.de}
% 2nd. author
\alignauthor
Cristiano Bertolini\\
       \affaddr{Informatics Center}\\
       \affaddr{Federal University of Pernambuco, Brazil}\\
       \email{cbertolini@cin.ufpe.br}
% 3rd. author
\alignauthor
Martin Sch\"af\\
       \affaddr{International Institute for Software Technology}\\
       \affaddr{United Nations University, Macau}\\
       \email{schaef@iist.unu.edu}
\and  % use '\and' if you need 'another row' of author names
% 4th. author
\alignauthor
Ishan Banerjee\\
       \affaddr{Dept. of Computer Science}\\
       \affaddr{University of Maryland}\\
       \email{ishan@cs.umd.edu}
% 5th. author
\alignauthor
Atif M. Memon\\
       \affaddr{Dept. of Computer Science}\\
       \affaddr{University of Maryland}\\
       \email{atif@cs.umd.edu}
}

\maketitle

\begin{abstract}
Graphical user interfaces (GUIs) encode, as event sequences, 
potentially unbounded ways to interact with software.
During testing it becomes necessary to effectively sample the GUI's event space. 
Ideally, for increasing the efficiency and effectiveness of GUI testing, one
would like to sample the GUI's event space by only generating
sequences that (1) are allowed by the GUI's structure, and (2) chain together
only those events that have data dependencies between their event handlers. We
propose a new model, called an event-dependency graph (EDG) of the GUI
that captures data dependencies between the code of event handlers. We
develop a mapping between an EDG and an existing black-box model of
the GUI's structure, called an event-flow graph (EFG). We automate the EDG
construction in a tool that analyzes the bytecode of each event handler. We evaluate our
``grey-box'' approach using four open-source applications and compare it with
the EFG approach. Our results show that using the EDG
reduces the number of event sequences with respect to the EFG, while still
achieving at least the same coverage. Furthermore, we are able to detect 2 new
bugs in the subject applications.
\end{abstract}




\section{Introduction}
\label{sec:introduction}

A particular challenge for system testing of software applications that have a
graphical user interface (GUI) front-end is that the total number of all
possible sequences of \emph{user actions} is prohibitively large (in principle,
possibly infinite), even for relatively small applications. A reasonably sized
and effective sample needs to be selected for testing. GUI testing, i.e., system
testing the software through its GUI is important, because most of today's
software applications provide services to end-users via a GUI.

Each user interaction, e.g., pressing a key on the keyboard or clicking a mouse
button, triggers an \emph{event} in the application. An application responds to
an event by executing a piece of code called the \emph{event handler} associated
with the event. In GUI testing, a \emph{sequence of events} is an integral part
of a GUI \emph{test case}. In particular, a GUI test case consist of (1) a
precondition that must hold before executing a sequence of events; (2) the
actual sequence of events to be executed; (3) possible input-data to the GUI;
and (4) the expected results of the test case (the oracle).

There has been extensive recent work on developing automated model-based GUI
testing techniques. Current techniques (e.g.,
\cite{DBLP:conf/icst/BertoliniPdM09, Silva:2010, Paiva2010,
BryceSampathMemonTSE2011, YuanCohenMemonTSE2011, YuanMemonIST2010,
icse-2009-jkkmptv,bertolini2010}) use a \emph{black-box} approach to generate
test cases. Further, they use a graph-based model to represent the possible
sequences of events with the GUI. Each node in these graph-based models
represent an \emph{event}, which is an interaction with one \emph{widget} (e.g.,
selecting an element in a listbox). A path in this graph-based model corresponds
to a sequence of events with the GUI; this sequence is used in the GUI test
case.

% Typically there is an enormous number of paths in these graphs; current
% techniques sample from these paths using directives supplied manually by a
% tester; or by limiting event sequences to very short lengths (1 or
% 2)~\cite{AtifTSE.2010.12}. \Fix{it is also our limitation, right? maybe we can
% remove this paragraph\ldots}

In this paper, we propose and evaluate a \emph{grey-box}~\cite{kicillof2007}
approach for automated GUI testing. The underlying mechanism for the grey-box
approach is a new \emph{event-dependency graph} (EDG) model that captures data
dependencies between event-handlers in the GUI code. More specifically, an EDG
is a weighted directed graph in which each node represents an event in the GUI.
An edge from the node representing event $e_1$ to a node representing event
$e_2$ shows that there is a \emph{data dependency} from $e_1$'s event handler to
$e_2$'s event handler. The weight of the edge represents the number of fields
that flow from $e_1$'s event handler to $e_2$'s event handler. \AbstractTCs{}
are generated by using a minimax search~\cite{osborne-rubinstein} on the EDG. An
\abstractTC{} is a path through the EDG. Because of the nature of the EDG model,
these \abstractTC{} chain together only those events that have data dependencies
between their event handlers. Further, an \abstractTC{} does not necessarily
mean that their events are allowed one after the other by the GUI's structure.
For example, $e_1$ may be an event in the \texttt{MainWindow}, whereas event
$e_2$ may be in the \texttt{FileOpen} dialog. An intermediate event that opens
the \texttt{FileOpen} dialog is needed before $e_2$. Hence \abstractTCs{}, which
are paths in the EDG, may not be executable, which is why we called them
``abstract'' event sequences above. To convert \abstractTCs{} into
``executable'' event sequences, a mapping is maintained between the EDG and the
GUI's workflow, represented using an existing \emph{event-flow graph} (EFG)
black-box model of the GUI~\cite{Memon07}. After applying the mapping, we obtain
event sequences that (1) are allowed by the GUI's structure, and (2) chain
together only those events that have data dependencies between their event
handlers. By embedding these \execTCs{} into GUI test cases, a compact test
suite is formed, which efficiently samples GUI event space.

We evaluate the grey-box approach on four open-source applications:
\emph{TerpWord}, \emph{Rachota}, \emph{FreeMind} and \emph{JabRef}. The results show a dramatic
increase in the efficiency of the event sequence generation and execution.
Further, one new bug in Rachota, and one bug in JabRef is revealed.

The paper is organized as follows: Section~\ref{sec:blackbox} provides the
background of model-based GUI testing using a black-box approach.
Section~\ref{sec:greybox} introduces our grey-box GUI testing approach, which
incorporates an event-flow graph (EFG) and an event-dependency graph (EDG) to
generate efficient event sequences. Section~\ref{sec:impl} provides an overview
of the implementation, which we use to evaluate the approach
(Sections~\ref{sec:exp} through \ref{sec:discussion}). Section~\ref{sec:related}
summarizes the related work, and finally, Section~\ref{sec:conclusion} presents
the conclusions and future work.

\section{Background}
\label{sec:blackbox}

When testing a system through its GUI, only a finite set of user interactions
can be tested. The choice of this set is vital to the success of the testing
procedure. A common way to sample the possibly infinite set of sequences is to
use a graph-based model of the GUI, called \textit{event-flow graph} (EFG).

An event-flow graph, $EFG = \langle \eventset, \initset, \efgedge \rangle$, for
an application is a directed graph. Each node $\event\in\eventset$ is an event
in the GUI. An event is a response of the system to a user interaction (a click
on a button triggers an \texttt{onClick} event). Each event in
$\initset\subseteq\eventset$ is an initial event which can be executed directly
after the application launched. An edge $(\event,\event') \in \efgedge$ between
to events $\event, \event' \in \eventset$ states that the event $\event'$ can be
executed immediately after the event $\event$. Conversely, if there is no edge
between events $\event, \event'$ then event $\event'$ cannot be executed
immediately after event $\event$. This may be owing to structural
characteristics of the GUI. For example, executing $\event$ may close the window
containing $\event'$. The EFG can be obtained automatically from the application
using a \emph{GUI Ripper}~\cite{MemonWCRE2003}. Section~\ref{sec:impl-efg}
outlines the construction of the EFG, its benefits and limitations.

Figure~\ref{fig:example-efg}(a) shows the GUI of an example application. The
\texttt{MainWindow} appears when the application is launched. A modal dialog
\texttt{Dialog} appears when the button $e_3$ is clicked. It is closed when the
button $e_4$ is clicked.

\input{pics/example-efg}

Figure~\ref{fig:example-efg}(b) shows the corresponding EFG of the example
application, which consists of 4 events ($e_1$ to $e_4$), where the events
$e_1$, $e_2$, $e_3$ represent initial events. The execution event $e_3$ opens
the modal dialog, s.t. $e_4$ becomes accessible. The event $e_4$ closes the
\texttt{Dialog} and thus, after $e_4$ is executed, it becomes inaccessible
again.

An \textit{event sequence} in an EFG is a sequence of events which represents a
sequence of user interactions with the GUI. An \execTC{} $\tc = \event_0, \ldots,
\event_n$ is a sequence of events which starts with an initial event
$\event_0\in\initset$.

\begin{definition}
\label{def:testcase}
Given an event-flow graph $EFG = \langle \eventset,\initset, \efgedge \rangle$.
An \execTC{} is a sequence of events $\tc = \event_0, \ldots,
\event_n$, such that $\event_0\in\initset$ and $(\event_i,\event_{i+1}) \in
\efgedge$ for all $0\leq i < n$.
\end{definition}

From the $EFG$, sequences of events of a particular length are sampled. For
instance, using a sequence length of $1$ leads to the following event sequences:
$s_1 = \langle {\bf e_1} \rangle$, $s_2 = \langle {\bf e_2} \rangle$, $s_3 =
\langle {\bf e_3} \rangle$, and $s_4 = \langle e_3, {\bf e_4} \rangle$. Note
that sequence $s_4$ has length 2. This is because $e_4$ cannot be tested with a
sequence of length~1, therefore additional \emph{reaching steps} are introduced
to connect $e_4$ to an initial event of the $EFG$.

Although event sequences of length~1 provide a compact set, it is certainly not
sufficient for bug detection, since pairs and triples of events are not
considered. However, increasing the length of the event sequences does not
scale, as the number of generated sequences grows exponentially
(Table~\ref{tab:example-efg} shows all event sequences generated with a length
of $2$ for the EFG in Figure~\ref{fig:example-efg}(b)). That is, a better
technique for sampling the $EFG$ is needed in order to generate event sequences
with a reasonable length. In the following we present a technique to
efficiently generate a compact set of relevant event sequences of arbitrary length.

\begin{table}[h]
\begin{tabular}{llll}
$s_1 = \langle {\bf e_1, e_1} \rangle$ &
$s_5 = \langle {\bf e_2, e_2} \rangle$ &
$s_9 = \langle e_3, {\bf e_4, e_2} \rangle$
\\

$s_2 = \langle {\bf e_1, e_2} \rangle$ &
$s_6 = \langle {\bf e_2, e_3} \rangle$ &
$s_{10} = \langle e_3, {\bf e_4, e_3} \rangle$
\\

$s_3 = \langle {\bf e_1, e_3} \rangle$ &
$s_7 = \langle {\bf e_3, e_4} \rangle$ &
\\

$s_4 = \langle {\bf e_2, e_1} \rangle$ &
$s_8 = \langle e_3, {\bf e_4, e_1} \rangle$ &
\\
\end{tabular}
\caption{Generated Event Sequences using an EFG Sequences Length of 2}
\label{tab:example-efg}
\end{table}

\section{Grey-box GUI Testing}
\label{sec:greybox}

An EFG is useful to generate feasible event sequences. However, when generating
longer event sequences the number sequences becomes prohibitively large and a
more sophisticated sampling strategy is needed.

To efficiently sample the event sequences generated from an EFG, we propose to
incorporate additional information from the source code of the event handlers.
Knowing which fields are modified and which are read upon the execution of an
event makes it possible to prioritize sequences of events where the event
handlers influence each other and to avoid those sequences, where events are
completely independent (e.g., a \texttt{Copy} and a \texttt{Help} button in a
word processor).

\begin{center}
\begin{minipage}{.41\textwidth}
\begin{lstlisting}[language=Java,caption=Java Snippet of the Example Event
Handlers.,label=lst:example-java]
class MainWindow {
	boolean enabled = true;
	String text = "Hello World";
	
	void e1() {
		enabled = false;
	}
	
	void e2() {
		text = text.toLowerCase();
	}
	
	void e3() {
		if ( enabled )
			openDialog(this);
		else
			Log.write(text);
	}
}

class Dialog {
	MainWindow mainWindow;
	
	void e4() {
		mainWindow.text = null;
		closeDialog();
	}
}
\end{lstlisting}
\end{minipage}
\end{center}

Listing~\ref{lst:example-java} shows the Java snippet of the example
application, especially of their 4 event handlers. The example application
consists of the classes \texttt{MainWindow} and \texttt{Dialog}, where
\texttt{MainWindow} contains three event handlers (\texttt{e1}, \texttt{e2}, and
\texttt{e3}), and \texttt{Dialog} the event handler \texttt{e4}. Event handler
\texttt{e1} sets the field \texttt{enabled} to \texttt{false}, and \texttt{e2}
converts the string of field \texttt{text} to lower case. In \texttt{e3}, the
field \texttt{enabled} is evaluated in a conditional. If \texttt{enabled} is
\texttt{true}, the dialog is opened and the current instance \texttt{this} of
\texttt{MainWindow} is passed to the dialog. If \texttt{enabled} is false, the
content of \texttt{text} is written to a log. Event handler \texttt{e4} sets the
field \texttt{text} of the current instance of \texttt{MainWindow} to
\texttt{null} and closes the dialog.

The execution of the event sequence $\langle e_3, e_4, e_2 \rangle$ throws a
\texttt{NullPointerException}, because the field \texttt{text} in \texttt{e2}
was set to \texttt{null} in \texttt{e4}. This example application is a
simplified version of a bug which we found in real world applications.

Without considering the application's source code, in the worst case, all
sequences of length 2 must be generated and executed to detect the bug. For our
example, this leads to 10 event sequences in total. When analyzing the source
code of an application, we observe that certain event handlers share a
\emph{data dependency}, which helps to prefer or to neglect certain events from
event sequence generation: Event \texttt{e1} writes field \texttt{enabled} which
is read in \texttt{e3}; \texttt{e4} writes field \texttt{text}, which is read in
\texttt{e2}. Further, there is no data dependency between $e_1$ and $e_2$.
To utilize these data dependencies for a more efficient event sequence
generation, we introduce a new graph-based mode called event-dependency graph
(EDG).

\subsection{Event-dependency Graph}
An event-dependency graph $EDG = \langle \eventset,\edgedge \rangle$ is a
directed graph where, like in the EFG, each node in $\eventset$ represents a GUI
event. Note that in contrast to the EFG, an EDG does not have initial events
since it represents data dependency and not control-flow. An edge $(\event,
\weight, \event')\in \edgedge$ is labeled with a weight $\weight$. The weight
$\weight \in \mathbb{N}^{+}$ indicates the data dependency between $\event$ and
$\event'$.

The edge value ($\weight$) is computed as follows: All fields which are
written in the event handler of $\event$ are collected in a set $W$. All
fields that are read in the event handler of $\event'$ are collected in a set
$R$. For each event handler, we recursively follow potential method calls,
collect these fields, and place them in set $W$ and $R$ respectively. The
edge from $\event$ to $\event'$ is labeled with the size of the intersection of
these set $|R\cap W|$.

A path $\pi = \event_i \ldots \event_j$ in the EDG represents a sequence of
events, where the execution of one event always changes fields which are read
by the succeeding event. However, it is not necessary that two events in
question can be executed consecutively in the GUI. The benefit of these
sequences is that the execution of one event might change relevant fields for
the execution of its successor and cause this one to execute other code fragments. 
This can lead to a higher code coverage and further reduce the amount
of code that is tested redundantly. Since the EDG has no initial events, and
succeeding events on a path in the EDG might not be directly executable in the
GUI, we refer to an EDG path as an \abstractTC{}.

\begin{definition}
\label{def:abstracttestcase}
Given an event-dependency graph $EDG = \langle \eventset,\edgedge \rangle$. An
\abstractTC{} is a sequence of events $\pi = \event_i, \ldots,
\event_j$, such that $(\event_k,\event_{k+1})\in\efgedge$ for all $i\leq k < j$.
\end{definition}

\LinesNumbered
\SetAlFnt{\small}
\begin{algorithm}[h]
\KwIn{$P$ : Program, \\
\quad\quad\quad $\langle \eventset, \initset,\efgedge \rangle $ : Event-flow graph}
\KwOut{$\langle \eventset',\edgedge \rangle$ : Event-dependency graph}
\SetKwFunction{getFieldsWritten}{getFieldsWritten}
\SetKwFunction{getFieldsRead}{getFieldsRead}
\DontPrintSemicolon
\Begin{
  $\eventset' = \eventset$ \;
  W = \{\}$, $R = \{\} \;
  \ForEach{($\event$ in $\eventset$)}{
    $W = \getFieldsWritten(e,P)$ \;
    \ForEach{($\event'$ in $\eventset$)}{
      $R = \getFieldsRead(e',P)$ \;
      \If{($(R \cap W) \neq \emptyset$)}{
      	$\weight = |R \cap W| $ \;
        $\edgedge = \edgedge \cup (\event, \weight, \event')$ \;
      }
    }
  }
}
\caption{Construction of the EDG.}
\label{alg:edg}
\end{algorithm}

Algorithm~\ref{alg:edg} shows how the EDG is constructed. The algorithm takes
the program $P$ and a corresponding event-flow graph $EFG$ as input, and returns
an event-dependency graph $EDG$. Since both EFG, and EDG refer to the same set
of events, we copy $\eventset$ to $\eventset'$ (line~2). Then, we iterate over
all pairs of events $\event,\event'$ (line~4).

We call the method \texttt{getFieldsWritten} which returns a set $W$ of all
fields that are written during the execution of the event handler of $\event$
(line~5). Then, we call the method \texttt{getFieldsRead} that returns a set
$R$ of all fields which are read during the execution of the event handler of
$\event'$ (line~7). If the intersection of $R$ and $W$ is not empty (line~8), we
add a new edge to the edge which is labeled with the size of the intersection
(line~10). Note that our algorithm does not create an edge between events if the
intersection of $R$ and $W$ is empty. In this case, there is no data dependency
between both events and thus, they are not directly connected (otherwise the EDG
would be fully connected).

\subsection{Event Sequence Generation}
Our event sequence generation is built out of two consecutive steps. First, we
select potentially interesting sequences of events, called \abstractTC{}, from
the EDG using Algorithm~\ref{alg:tcg}. Second, we use the \abstractTCs{} to
generate \execTCs{} from the EFG using Algorithm~\ref{alg:edg2efg}.

Algorithm~\ref{alg:tcg} takes an EDG and two parameters as input: $\tlen$ gives
the maximum length of the \abstractTC{} to be generated, and $\ttop$ gives the
maximum number for \abstractTCs{} to be generated for each event.
The algorithm returns a set $\pathset$ of \abstractTCs{}. These are later used
for generating \execTCs{}.

\LinesNumbered
\SetAlFnt{\small}
\begin{algorithm}[h]
\KwIn{$\langle \eventset,\edgedge \rangle$ : Event-dependency graph, \\
\quad\quad $\tlen$ : max length of \abstractTC{}, \\
\quad\quad $\ttop$ : max number of \abstractTCs{} per event}
\KwOut{$\pathset$ : set of \abstractTCs{}}
\SetKwFunction{bestsucc}{bestSucc}
\SetKw{post}{post}
\SetKw{from}{from}
\SetKw{pick}{pick}
\DontPrintSemicolon
\Begin{    
  Sequences of events $\pathset = \{\}$\;
  
  \ForEach{Event $\event \in \eventset$} {
    Sequences of events $\pathset' = \{\}$\;
    \While{$|\pathset'| < \ttop$} {
      Sequence of events $\pi = \event$ \;
      Event $\event'=\event$\;
      \While{$|\pi|<\tlen \wedge \post(\event')\neq \{\}$} {
        $\event' = \bestsucc(\event', \pathset)$ \;
        $\pi = \pi \bullet \event'$ \;
      }
      \lIf {$\pi\in\pathset$}{break}\;
        $\pathset' = \pathset' \cup \{\pi\}$\;
    }
    $\pathset = \pathset \cup \pathset'$\;
  }
  \Return $\pathset$ \;
}
\caption{Generating \abstractTCs{}.}
\label{alg:tcg}
\end{algorithm}

For each event $\event \in \eventset$, a new set $\pathset'$ of \abstractTCs{}
is created, which initially is empty (line~4). As long as the size of this set
is smaller than $\ttop$ (line~5), we add further \abstractTCs{} (line~6).
Each such \abstractTC{} $\pi$ initially contains only $\event$ (as we are
looking for sequences of events that start in $\event$). While the length of
$\pi$ is smaller than $\tlen$ (line~8), and the last event of $\pi$ still has
successors, the method \texttt{bestSucc} is finds the best possible successor
and adds it to the end of $\pi$ (line~10). The method \texttt{bestSucc} uses a
\emph{minimax} strategy to identify the best successor, unless this successor
leads us on a path which is already in $\pathset'$. In this case,
\texttt{bestSucc} returns the second best choice. We use the \emph{minimax}
strategy to minimize the selection of events with low dependencies.

The loop in line~5 terminates either, if it has collected $\ttop$ \abstractTCs{}
that start with the event $\event$ or, if the algorithm detects a path twice
(line~12). In that case, \texttt{bestSucc} cannot find a suitable path that has
not been visited so far.

For each \abstractTC{} in $\pathset$, we want to generate an \execTC{}. However,
the \abstractTCs{} are not necessarily executable, as consecutive events in the
EDG might have no direct connection in the EFG. Therefore, we use
Algorithm~\ref{alg:edg2efg} to find one EFG path for each of these
\abstractTCs{}, which starts in an initial event of the EFG. Note that the only
case, where such a path does not exist is, if the application is terminated
between the execution of two events. In that case, we split the sequence into
two sequences that later on are tested immediately after each other.

\LinesNumbered
\SetAlFnt{\small}
\begin{algorithm}[h]
\KwIn{$\langle \eventset,\initset, \efgedge \rangle$ : Event-flow graph, \\
\qquad\qquad\qquad\,$\Pi$ : set of \abstractTCs{}}
\KwOut{ $\testset$ : Set of \execTCs{}}
\SetKwFunction{shortpath}{shortestPath}
\SetKw{from}{from}
\SetKw{pick}{pick}
\DontPrintSemicolon
\Begin{
  Sequences of events $\testset = \{\}$\;
  
  \ForEach{Sequence $\event_i, \ldots, \event_j$ in $\Pi$} {
    \pick $\event_0$ \from $\initset$ \;
    Path $\tc = \shortpath(\event_0, \event_i)$ \;
    \For{$k = i$ \KwTo $j-1$} {
      $\tc = \tc \bullet \shortpath(\event_k, \event_{k+1})$ \;
    }
    $\testset = \testset \cup \{\tc\}$\;
  }
  \Return $\testset$ \;
}
\caption{Conversion from \abstractTCs{} to \execTCs{}.}
\label{alg:edg2efg}
\end{algorithm}

Algorithm~\ref{alg:edg2efg} takes an EFG and the set $\pathset$ of
\abstractTCs{} computed by Algorithm~\ref{alg:tcg} as input, and returns a set
of \execTCs, which are paths in the EFG and start in an initial event.
For each sequence of events $\event_i\ldots\event_j$ in $\pathset$ (line~3), a
path $\tc$ (line~5) is created. We pick the shortest path from an event
$\event_0 \in \initset$ to $\event_i$, and then iterate over the events in the
abstract event sequences and always add the shortest path between succeeding
events to $\tc$ (line~7). Then we add $\tc$ to the set $\testset$ (line~9).
Since paths in $\testset$ start in an initial event of the EFG, it can
immediately executed as a GUI event sequence.\\

Infeasible event sequences are only generated if the EFG is not complete (e.g.,
because it was generated automatically) or if the data dependency analysis is
imprecise. As these are implementation issues, we refer to
Section~\ref{sec:impl} for details.

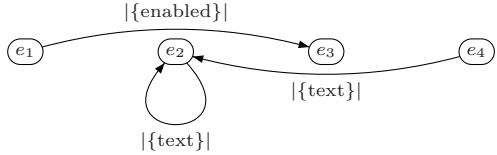
\begin{figure}[h]
  \center
  \begin{picture}(60,15)
	\scriptsize
	\gasset{Nadjust=wh,Nadjustdist=1}

	\node(e1)(0,10){$e_1$}
	\node(e2)(20,10){$e_2$}
	\node(e3)(40,10){$e_3$}
	\node(e4)(60,10){$e_4$}

	% e1
	\drawedge[curvedepth=3](e1,e3){|\{enabled\}|}

	% e2
	\drawloop[loopangle=-90](e2){|\{text\}|}

	% e4
	\drawedge[curvedepth=3](e4,e2){|\{text\}|}
  \end{picture}
  \caption{EDG of the Example Application.}
  \label{fig:example-edg}
\end{figure}

Figure~\ref{fig:example-edg} shows the EDG of our example application.
If we apply Algorithms~\ref{alg:tcg} with $\tlen = 2$ and $\ttop = \infty$,
Algorithm~\ref{alg:edg2efg} outputs the following \execTCs{}:
$e_1$ writes into $e_3$, which results in $s_1 = \langle {\bf e_1, e_3}
\rangle$. Since the field \texttt{text} is both read and written in $e_2$, $s_2
= \langle {\bf e_2, e_2} \rangle$ is generated. $e_3$ does not write into any
other event, and thus, is considered in a single event sequence $s_3 = \langle
{\bf e_3} \rangle$. Finally, $e_4$ writes into $e_2$, which leads to $s_4 =
\langle e_3, {\bf e_4, e_2} \rangle$. Because $e_4$ does not represent an
initial event, the intermediate event $e_3$ is inserted.

Note that it is not possible to combine EFG and an EDG into one graph-based
model: On the one hand, it is possible to label a directed edge $(e_1, e_2)$ in
the EFG with the weight of the data dependency (e.g., zero in case of a
non-dependency). On the other hand, a directed weighted edge $(e_3, e_4)$ has to
be added to the EFG, if a data dependency is detected. However, the added edge
may represent an event sequence, which is not allowed in the GUI.

\section{Implementation}
\label{sec:impl}

We integrate an implementation of the grey-box approach into
GUITAR\footnote{http://guitar.sourceforge.net/}, which is a open source,
model-based system for automated GUI testing. Figure~\ref{fig:guitar} presents
an overview of the GUITAR system. The grey-highlighted steps in the overview
emphasize our extensions made to the GUITAR system. Considering the grey-box
approach, testing an application using the GUITAR system consists of the
following steps:

\begin{figure}[h]
  \center
  \begin{picture}(40,40)
	\scriptsize
	\gasset{Nadjust=h,Nadjustdist=1,Nw=13,Nmr=0}
	\node[fillcolor=Black](gui)(20,35){\textcolor{White}{GUI}}
	\node(bytecode)(20,30){Bytecode}

	\gasset{Nadjust=wh,Nadjustdist=1,Nmr=1}
	\node[fillgray=0.8](ripping)(0,20){(1) GUI Ripper}
	\node(efg)(0,0){(2) EFG Construction}
	\node[fillgray=0.8](edg)(20,10){(3) EDG Construction}
	\node[fillgray=0.8](tcg)(40,0){(4) Event Sequence Generator}
	\node(replay)(40,20){(5) Replayer}

	\drawedge[curvedepth=-4](gui,ripping){}
	\drawedge[curvedepth=-1](ripping,efg){}
	\drawedge[curvedepth=1](efg,edg){}
	\drawedge[curvedepth=0](bytecode,edg){}
	\drawedge[curvedepth=-1](efg,tcg){}
	\drawedge[curvedepth=1](edg,tcg){}
	\drawedge[curvedepth=-1](tcg,replay){}
	\drawedge[curvedepth=-4](replay,gui){}

	\drawrect[dash={0.5}0](9,40,31,25)
	\nodelabel[ExtNL=y,NLangle=155,NLdist=0.9](gui){AUT}
  \end{picture}
  \caption{Overview of the GUITAR System, including the Grey-box Extensions.}
  \label{fig:guitar}
\end{figure}
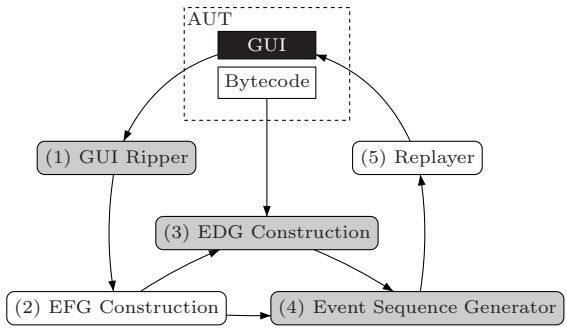

\subsection{GUI Ripper}
\label{sec:impl-ripper}

In the first step, the \emph{GUI Ripper} executes the AUT and records the GUI
structure. A GUI structure consists of widgets (e.g., windows, buttons, and text
fields) and their corresponding properties (e.g., enabled/disabled, height, and
width). While executing the AUT, the GUI ripper enumerates all widgets of the
main window using reflection, and stores the obtained information in the GUI
structure. For each found widget (e.g., a button), GUI ripper triggers the
assigned event (i.e., a button click). For instance, if the click on the button
opens a new window, GUI ripper continues to record the GUI structure of that
recently opened window and so on. The process stops, if all found windows have
been explored. Since each GUI represents a hierarchical structure, a depth-first
search is performed on the AUT's GUI. For the grey-box approach, we enhanced the
GUI ripper, such that, for each widget the event handlers assigned to
this widget are additionally stored in the GUI structure. This information is
needed during the analysis of the bytecode which is performed as a part of the
EDG construction.

\subsection{EFG Construction}
\label{sec:impl-efg}

The GUI structure recorded by the Ripper serves as input to the \emph{EFG
Construction}, which automatically constructs the EFG that is used for the test
case generation. While the GUI structure contains information about widgets and
their properties, the EFG represents an abstract view which only contains the
events and their following events. The EFG construction iterates over all
windows in the GUI structure and creates a single EFG for each window. Later,
these single EFGs are connected to one EFG representing the entire application.
For each window in the GUI structure, the EFG construction creates an event for
the window itself and for each containing widget. Then, the EFG construction
connects events of the window based on their widget properties. For instance, if
an event $e_1$ represents a window, and an event $e_2$ an enabled button in this
window, then an edge from $e_1$ to $e_2$ is created in the EFG. Assume, that
$e_2$ is associated to a disabled button, then no edge between $e_1$ and $e_2$
is created, because the event can not be triggered if the window appears.
For each window an EFG is created and the event which opens/accesses other
window is connected with all initial events of this other window. The details of
the EFG construction can be found in~\cite{MemonWCRE2003}.

\subsubsection{Short Assessment of the EFG Construction}
Since the GUI ripper performs a dynamic analysis of the GUI, it cannot be
guaranteed to find all widgets of the AUT~\cite{Memon07}. For instance, the AUT
itself might be hostile or even faulty, e.g., if the GUI opens a new window in
the background, the GUI ripper will not be able to find it, and thus, it cannot
be considered during EFG construction. Further, the fact if a widget is enabled
or disabled during ripping may strongly depends on the environment (e.g., user
settings). These problems tend to be of technical nature and their severity
might differ depending on the used platform.

Instead of executing the AUT in GUI ripping, it is in general possible to create
the GUI structure and the EFG respectively via static analysis. However, a
static analysis technique must be tailored to comprehend how a GUI is created.
While there exist different code styles for creating GUI's, a static technique
might find its limitations even if a GUI is defined outside the source code of
the application, e.g., in XML files.

Note that the EFG of an AUT is not complete and represents an approximation of
the AUT's event-flow. It cannot be guaranteed that a path in the constructed EFG
is actually executable on the AUT's GUI. For instance, if a click on a button
changes the entire parent window (e.g., removing or adding widgets), then the
GUI ripper and the EFG construction respectively does not recognize these
changes made to the GUI. A test engineer has to improve manually the EFG
according to the actual behavior of the AUT.

\subsection{EDG Construction}
\label{sec:impl-edg}

In order to construct an EDG, we perform a shallow
bytecode~\cite{Lindholm:1999:JVM:553607} analysis of the AUT to obtain data
dependency between events. In particular, the bytecode analysis records, which
fields are read and written by each event handler, that is, the functions
\texttt{getFieldsRead} and \texttt{getFieldsWritten} in Algorithm~\ref{alg:edg}.
Hence, the Java bytecode and the constructed EFG of the AUT serve as input to
the \emph{EDG Construction}. For our bytecode analysis, we use the the ASM
framework\footnote{http://asm.ow2.org/}. Other frameworks such as
Soot\footnote{http://www.sable.mcgill.ca/soot/} could be used equally well.

\subsubsection{Bytecode Analysis}

Listing~\ref{lst:example-bytecode} shows the bytecode of the event handler
\texttt{e1} and \texttt{e4} from the example application in
Listing~\ref{lst:example-java}. In bytecode, fields are read by the instruction
\texttt{GETFIELD}, and written by \texttt{PUTFIELD}. Further, methods are called
using the \texttt{INVOKE}\footnote{\texttt{INVOKEVIRTUAL},
\texttt{INVOKESTATIC}, \texttt{INVOKESPECIAL}} instruction. In line~2, a
constant value of $0$ is first pushed to the stack, and then assigned to field
\texttt{enabled} in line~3. In line~6 and 7 respectively, field
\texttt{mainWindow} and a constant value of \texttt{null} are pushed to the
stack. Field \texttt{text} of \texttt{mainWindow} is then assigned with the
value \texttt{null}. Finally in line~9, method \texttt{closeDialog} of is
called.

\begin{center}
\begin{minipage}{.41\textwidth}
\begin{lstlisting}[language=JVMIS,caption=Bytecode Snippet of the Example Event
Handlers.,label=lst:example-bytecode,basicstyle=\ttfamily\scriptsize]
void e1()V
 ICONST_0
 PUTFIELD MainWindow.enabled : Z

void e4()V
 GETFIELD Dialog.mainWindow : LMainWindow;
 ACONST_NULL
 PUTFIELD MainWindow.text : Ljava/lang/String;
 INVOKEVIRTUAL Dialog.closeDialog()V
\end{lstlisting}
\end{minipage}
\end{center}

\begin{figure}
\center
\begin{picture}(60,20)
	\scriptsize
	\gasset{Nadjust=wh,Nadjustdist=1,Nmr=0}

	\node(c)(0,20){Class}
	\node(m)(50,10){Method}
	\node(f)(0,0){Field}

	\drawedge[AHnb=0](c,f){contains}
	\drawedge[AHnb=0, curvedepth=3](c,m){contains}
	\drawloop[AHnb=0,loopangle=0](m){calls}
	\drawedge[AHnb=0,curvedepth=-3](m,f){reads}
	\drawedge[AHnb=0,curvedepth=3](m,f){writes}

	% class contains field
	\nodelabel[ExtNL=y,NLangle=-70,NLdist=1.0](c){1}
	\nodelabel[ExtNL=y,NLangle=70,NLdist=1.0](f){*}
	
	% class contains method
	\nodelabel[ExtNL=y,NLangle=20,NLdist=1.0](c){1}
	\nodelabel[ExtNL=y,NLangle=-215,NLdist=0.7](m){*}

	% method calls method
	\nodelabel[ExtNL=y,NLangle=55,NLdist=0.7](m){*}
	\nodelabel[ExtNL=y,NLangle=-55,NLdist=1.0](m){*}

	% method reads field
	\nodelabel[ExtNL=y,NLangle=190,NLdist=1.0](m){*}
	\nodelabel[ExtNL=y,NLangle=10,NLdist=1.0](f){*}
	
	% method writes field
	\nodelabel[ExtNL=y,NLangle=215,NLdist=1.0](m){*}
	\nodelabel[ExtNL=y,NLangle=-15,NLdist=1.0](f){*}
\end{picture}
\caption{Simplified ER Model of the ClassDB.}
\label{fig:classdb}
\end{figure}
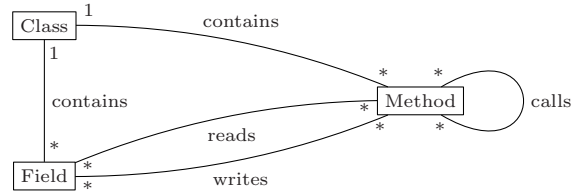

The EDG construction is preceded by one step: the creation of a class database
(ClassDB). The ClassDB models the dependencies between fields, methods and
classes of the AUT. During EDG construction, a request to the ClassDB determines
the data dependency of two given event handlers. Figure~\ref{fig:classdb} shows
the ER model of the ClassDB.

In order to build a ClassDB, the bytecode analysis starts with visiting all
classes of the AUT, since classes contain both methods and fields. In our
implementation, it is possible to provide a \emph{scope} (a set of JAR archives)
to restrict the set of classes to be analyzed. For instance, only application
classes are supposed to analyze and third-party libraries are discarded. Each
class is stored in table \texttt{Class} of the ClassDB and is identified by its
fully-qualified name, to avoid collisions if a certain class name is multiply
used.

Then, the bytecode analysis visits all methods of each class. Note that it is
important to inspect all methods, and not only those which are declared as event
handlers. Moreover, it is necessary to follow all methods calls in each method,
which can be detected by visiting the \texttt{INVOKE} instructions of the
bytecode. For instance, method \texttt{e4} in Listing~\ref{lst:example-bytecode}
calls method \texttt{closeDialog}, which may write further fields. Thus, there
exist a recursive relationship \emph{calls} between methods. Each method is
stored in table \texttt{Method} in the ClassDB and is associated to its class.

For each method, the bytecode analysis fetches all fields that are read and
written. This is can be detected by visiting the \texttt{GETSTATIC} and
\texttt{PUTSTATIC} instructions of the bytecode. Read and written fields are
stored in table \texttt{Field}, where each field is associated to its method.

Once all classes, methods and fields are visited and mapped in the ClassDB,
Algorithm~\ref{alg:edg} uses this information to construct the EDG. For
instance, if the algorithm requests the \texttt{getFieldsRead} and
\texttt{getFieldsWritten} for a certain event $e$, the ClassDB aggregates all
called method within the event handler of $e$. For each called method, and for
the event handler itself, the read and written fields are collected and returned
to the EDG construction. In this way, a possible data dependency between events
is captured. Further, due to this shallow analysis of the bytecode, the
computation time for building the ClassDB is low, even for big applications.\\

\subsubsection{Short Assessment of the EDG Construction}

Java distinguishes between instance fields and class fields, which are treated
the same way in our bytecode analysis. That is, not only class fields are mapped
to a certain class in the ClassDB, but also instance fields. Moreover, instance
fields are not mapped to their objects. Further, the bytecode analysis does not
distinguish between calls of instance methods and class methods and thus, is
not reliable regarding polymorphism.

The bytecode analysis does not consider potential aliasing of fields or
potentially infeasible control-flow. Hence, the resulting EDG is only an
approximation of the actual data dependencies between fields. However, we are
interested in prioritizing events, so a \emph{cheap} bytecode analysis in terms
of computation time is sufficient, while leaving room for further in-depth
analyses.

\subsection{Event Sequence Generator}
\label{sec:impl-tcg}

The Event Sequence Generator takes as input an EFG and an EDG from the
application. In this step, the Algorithms~\ref{alg:tcg} and \ref{alg:edg2efg}
are applied. The output is a set of \execTCs{}, where each \execTC{} is embedded
into one GUI test case.

\subsection{Replayer}
\label{sec:impl-replayer}

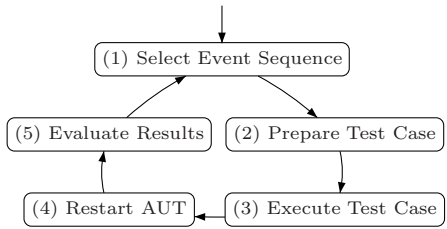
\begin{figure}[h]
  \center
  \begin{picture}(60,25)
	\scriptsize
	\gasset{Nadjust=wh,Nadjustdist=1,Nmr=1}

	\node[Nmarks=i,iangle=90](select)(30,20){(1) Select Event Sequence}
	\node(prepare)(45,10){(2) Prepare Test Case}
	\node(execute)(45,0){(3) Execute Test Case}
	\node(restart)(15,0){(4) Restart AUT}
	\node(evaluate)(15,10){(5) Evaluate Results}

	\drawedge[curvedepth=1](select,prepare){}
	\drawedge[curvedepth=1](prepare,execute){}
	\drawedge[curvedepth=1](execute,restart){}
	\drawedge[curvedepth=1](restart,evaluate){}
	\drawedge[curvedepth=1](evaluate,select){}
  \end{picture}
  \caption{Overview of the Replayer Process.}
  \label{fig:replayer}
\end{figure}

The Replayer is responsible for executing GUI test cases. A test case is
considered as a precondition, an \execTC{}, input-data and an oracle.
Figure~\ref{fig:replayer} presents an overview of the Replayer process. It
consists of the following steps: (1) it selects an \execTC{}; (2) it prepares a
test case, which ensures that the precondition of the test case holds; (3) it
executes the test case on the AUT, which performs the \execTC{}; (4) it restarts
the AUT, which covers the events \texttt{exit} and \texttt{launch} of the AUT;
(5) it evaluates, whether the test case has failed or passed.

\section{Experiment}
\label{sec:exp}

We compare our grey-box approach with the black-box approach by studying
efficiency and effectiveness. \emph{Efficiency} is considered as the computation
time for generating the \abstractTCs (in minutes) and the time for test case
execution (in hours). \emph{Effectiveness} is considered as the line and branch
coverage (in percentage). We define two research questions. \textbf{Q1}: Is the
grey-box approach efficient in terms of mean time to execute the test cases? And
\textbf{Q2}: Is the grey-box approach effective in terms of mean code-coverage?

\subsection{Setup of the Experiment}
\label{sec:exp-setup}

We evaluate the grey-box approach using four Java-based open source
applications: \emph{TerpWord 4.0} is a word processor, \emph{Rachota 2.3}
is a time recording system. \emph{FreeMind 0.9.0} creates mind maps, and
\emph{JabRef 2.7} manages bibliographic references. It is important to observe
that we use stable versions of all applications where bugs are rarely found. We choose these applications to
consider both small and large applications (in terms of \# of classes), and to
cover different code styles. Table~\ref{tab:exp-setup} shows some relevant
statistics of the Applications Under Test (AUTs): the number of lines of code
(\textbf{LOC}), number of classes (\textbf{Classes}), number of GUI events
(\textbf{Events}), number of edges in the EFG (\textbf{EFG edges}), and number
of edges in the EDG (\textbf{EDG edges}).

\begin{table}[h]
\scriptsize
\begin{center}
    \begin{tabular}{|l|r|r|r|r|}
        \hline
        ~& \textbf{TerpWord} & \textbf{Rachota} & \textbf{FreeMind} & \textbf{JabRef} \\
        \hline \hline \textbf{LOC} & 6,842 & 13,750 & 40,922 & 68,468 \\
        \hline
        \textbf{Classes}   & 215 & 468 & 1,362 & 4,027 \\
        \hline
        \textbf{Events}    & 159 & 154 & 959  & 776 \\
        \hline 
        \textbf{EFG edges} & 4,229  & 1,493 & 105,986 & 100,211 \\
        \hline 
        \textbf{EDG edges} &  4,100 & 2,172 & 25,248  & 10,034 \\
        \hline
    \end{tabular}
	\caption{Experiment setup.}
	\label{tab:exp-setup}
	\end{center}
\end{table}

Table~\ref{tab:exp-conf} shows six different configuration used to test the four
applications. For brevity of exposure we use identifiers (\textbf{ID}) to refer
to these configurations.

The black-box approach presented in~\cite{Memon07} is used in Configuration A,
B, and C. These configurations are the baseline of our experiment.
Configuration A generates one event sequence (length = 1) for each event in the
EFG. Configuration B (length = 2) generates event sequences for each \emph{pair}
of events $(e_i, e_j)$, that have a direct connection in the EFG.
Configuration C (length = 3) generates event sequences for each \emph{triple}
of events $(e_i, e_j, e_k)$, where $\{e_i, e_j\}$ and $\{e_j, e_k\}$ are direct
neighbors in the EFG.

The grey-box approach is used in Configuration D, E. Configuration D considers
\abstractTCs{} of length 2 and does not limit the number of \abstractTCs{}
generated per event. Configurations E and F have \abstractTCs{} of length 3.
Here, the number of generated \abstractTCs{} is limited to 50 and 100
respectively. This is because each event in the applications \emph{TerpWord} and
\emph{FreeMind} has about 25 dependent events. That is, the number of
\abstractTCs{} of this size is already large. In particular, we are interested
in knowing, if doubling the number of generated \abstractTCs{} from 50 to 100
will have a significant impact on the coverage.

\begin{table}
\scriptsize
	\begin{center}
		\begin{tabular}{|c|l||c|l|}
			\hline \hline
			\multicolumn{2}{|c||}{\textbf{Black-box Approach}} &
			\multicolumn{2}{|c|}{\textbf{Grey-box Approach}}\\
			\hline
			\textbf{ID} & \textbf{Configuration} & \textbf{ID} & \textbf{Configuration}
			\\
			\hline \hline
			\textbf{A} & \tlen = 1 & \textbf{D} & \tlen~ = 2\\
			\hline
			\textbf{B} & \tlen = 2 & \textbf{E} & \tlen~ = 3; \ttop~ = 50\\
			\hline
			\textbf{C} & \tlen = 3 & \textbf{F} & \tlen~ = 3; \ttop~ = 100 \\
			\hline 
		\end{tabular}
		\caption{Experiment configurations.}
		\label{tab:exp-conf}
	\end{center}
\end{table}

Each generated \execTC{} is embedded in a GUI test case, which is executed by
the Replayer. The precondition of each test case states, that all user-settings
have to be deleted before performing the event sequence on the AUT. As an
oracle, a crash monitor is used, which records any exception found during the
test case execution.

The test cases are executed on 10 Linux machines with a 4 x 2.0 GHz CPU, 4 GB
RAM, 500 GB HDD. The experiment was executed two times with the same setup
(e.g., the same seed for input-data) to ensure that the obtained results are
reproducible. The total number of test cases executed is 6,089,403.

\subsection{Experimental Results}
\label{sec:exp-results}

Table~\ref{tab:exp-results} shows a summary of the experimental results. For
each configuration we report the number of event sequences (\emph{\# es}),
broken event sequences (\emph{\# broken es}), total generation time (\emph{gen t
(m)}), generation time per event sequence (\emph{gen t per es (s)}), total
execution time (\emph{exec t (h)}), execution time per test case (\emph{exec t
per tc (s)}), line coverage (\emph{line cov.}) and branch coverage (\emph{branch
cov.}). The event sequence generation time is expressed in minutes and the test
case execution time in hours. The generation time per event sequence and
execution time per test case are expressed in seconds.

\begin{table}[!h]
\scriptsize
	\begin{center}
		\begin{tabular}{|l|r|r|r|r|}
			\hline
			& TerpWord & Rachota & FreeMind & JabRef \\
			\hline \hline
			\multicolumn{5}{|c|}{\textbf{Configuration A (Black-Box Approach)}} \\
			\hline
			\# es & 159 & 154 & 959 & 776 \\
			\hline
			\# broken es & 0 & 0 & 5 & 5 \\
			\hline
			gen t (m) & 0.3 & 0.28 & 1.10 & 1.08 \\
			\hline
			gen t per es (s) & 0.12 & 0.12 & 0.12 & 0.12 \\
			\hline
			exec t (h) & 0.50 & 0.58 & 4.58 & 4.22 \\
			\hline
			exec t per tc (s) & 12 & 15 & 30 & 28 \\
			\hline
			line cov. (\%) & 41 & 60 & 50 & 51 \\
			\hline
			branch cov. (\%) & 22 & 31 & 36 & 22 \\
			\hline \hline
			\multicolumn{5}{|c|}{\textbf{Configuration B (Black-Box Approach)}} \\
			\hline
			\# es & 3,307 & 1,310 & 11,396 & 43,017 \\
			\hline
			\# broken es & 0 & 0 & 57 & 258 \\
			\hline
			gen t (m) & 6.62 & 2.62 & 24.68 & 93.2 \\
			\hline
			gen t per es (s) & 0.12 & 0.12 & 0.13 & 0.13 \\
			\hline
			exec t (h) & 11.94 & 5.82 & 98.13 & 358.48 \\
			\hline
			exec t per tc (s) & 13 & 16 & 31 & 30 \\
			\hline
			line cov. (\%) & 55 & 61 & 53 & 54 \\
			\hline
			branch cov. (\%) & 36 & 34 & 37 & 26 \\
			\hline \hline
			\multicolumn{5}{|c|}{\textbf{Configuration C (Black-Box Approach)}} \\
			\hline
			\# es & 79,949 & 20,221 & 489,250 & 5,360,366 \\
			\hline
			\# broken es & 0 & 0 & 2,446 & 32,162 \\
			\hline
			gen t (m) & 159.90 & 40.44 & 1,223.13 & 15,187.70 \\
			\hline
			gen t per es (s) & 0.12 & 0.12 & 0.15 & 0.17 \\
			\hline
			exec t (h) & 310.92 & 95.49 & 4,348.89 & 44,669.72 \\
			\hline
			exec t per tc (s) & 14 & 17 & 32 & 30 \\
			\hline
			line cov. (\%) & 55 & 62 & 53 &  55 \\
			\hline
			branch cov. (\%) & 36 & 36 & 38 & 27 \\
			\hline \hline
			\multicolumn{5}{|c|}{\textbf{Configuration D (Grey-Box Approach)}} \\
			\hline
			\# es & 2,695 & 1,407 & 9,944 & 5,860 \\
			\hline
			\# broken es & 0 & 0 & 63 & 83 \\
			\hline
			gen t (m) & 7.63 & 4.22 & 43.08 & 20.52 \\
			\hline
			gen t per es (s) & 0.17 & 0.18 & 0.26 & 0.21 \\
			\hline
			exec t (h) & 9.73 & 6.25 & 88.39 & 48.83 \\
			\hline
			exec t per tc (s) & 13 & 16 & 32 & 30 \\
			\hline
			line cov. (\%) & 55 & 62 & 53 & 54 \\
			\hline
			branch cov. (\%) & 36 & 36 & 37 & 26 \\
			\hline \hline
			\multicolumn{5}{|c|}{\textbf{Configuration E (Grey-Box Approach)}} \\
			\hline
			\# es & 2,068 & 1,781 & 7,113 & 9,595 \\
			\hline
			\# broken es & 0 & 0 & 45 & 135 \\
			\hline
			gen t (m) & 6.55 & 5.93 & 35.57 & 36.78 \\
			\hline
			gen t per es (s) & 0.19 & 0.2 & 0.3 & 0.23 \\
			\hline
			exec t (h) & 9.19 & 8.91 & 65.20 & 90.62 \\
			\hline
			exec t per tc (s) & 16 & 18 & 33 & 34 \\
			\hline
			line cov. (\%) & 47 & 62 & 53 & 55 \\
			\hline
			branch cov. (\%) & 26 & 36 & 38 & 27 \\
			\hline \hline
			\multicolumn{5}{|c|}{\textbf{Configuration F (Grey-Box Approach)}} \\
			\hline
			\# es & 4,036 & 3,307 & 12,904 & 18,497 \\
			\hline
			\# broken es & 0 & 0 & 81 & 261 \\
			\hline
			gen t (m) & 13.45 & 11.57 & 68.82 & 77.07 \\
			\hline
			gen t per es (s) & 0.2 & 0.21 & 0.32 & 0.25 \\
			\hline
			exec t (h) & 17.94 & 17.45 & 118.29 & 179.83 \\
			\hline
			exec t per tc (s) & 16 & 19 & 33 & 35 \\
			\hline
			line cov. (\%) & 55 & 62 & 53 & 55 \\
			\hline
			branch cov. (\%) & 36 & 36 & 38 & 27 \\
			\hline
		\end{tabular}
		\caption{Results of the Experiment.}
		\label{tab:exp-results}
	\end{center}
\end{table}

\textbf{Observation 1:} Configuration A has the smallest number of event
sequences and the lowest coverage. The number of event sequences corresponds to
the number of events, which means that each event is only tested once. However,
we believe that this approach is useful for smoke
tests~\cite{Memon:2005:SFE:1100866.1100984}, since the generation and execution
time is also the lowest amongst all the configurations.

\textbf{Observation 2:} As expected, Configuration D is significantly more
efficient than Configuration B on the applications TerpWord, FreeMind, and
JabRef. For these applications, both configurations have the same line and
branch coverage. However, Configuration D uses significantly fewer event
sequences and consumes less time than Configuration B.
Thus, the grey-box approach generates more \emph{efficient} event sequences than
the black-box approach for these three applications.

\textbf{Observation 3:} For Rachota, Configuration D attains a higher coverage
than Configuration B. However, more event sequences are generated in
Configuration D, and the execution consumes more time. This is likely owing to
the fact that in Rachota, the EDG has significantly more edges than the EFG.

\textbf{Observation 4:} The number of generated event sequences of JabRef in
Configuration C exceeds 5 million. Comparing to Configuration B, the obtained
coverage for JabRef is disappointing with respect to the generation and
execution time. On the other hand, Configuration E and F are significantly more
\emph{efficient} on application Rachota, FreeMind, and JabRef, since fewer event
sequences are generated and executed while preserving the same line and branch
coverage.

\textbf{Observation 5:} For TerpWord, Configuration E attains a lower line and
branch coverage than for Configuration D and F. Thus, the parameter $\ttop$
influences the quality of the selected event sequences. In TerpWord, there are a
few events which have more than 50 dependent events. So, setting $\ttop = 50$
might not be effective enough. The Configuration F achieves more coverage when
setting $\ttop = 100$.

\textbf{Observation 6:} For the grey-box Configurations D, E and F, increasing
the length of the \abstractTCs{} does not significantly improve the coverage.

\textbf{Observation 7:} The number of \emph{broken event sequences} is
relatively low comparing to the total number of event sequences; they ranges
between 0,5\% and 1,4\%. Broken event sequences are sequences sampled from the
EFG, but could not be executed due to the limitations described in
Section~\ref{sec:impl-efg}. 

\textbf{Observation 8:} The experiment found 3 different bugs: The first bug is
found in JabRef with Configuration B and Configuration D.
A \texttt{NullPointerException} is thrown if the user clicks \texttt{Options},
\texttt{Manage custom imports}, \texttt{Add from fol\-der}, \texttt{Cancel}. The
bug is found in the black-box and in the grey-box approach using a sequence
length of 2.

\textbf{Observation 9:} The second bug was found in JabRef with Configuration D.
The following sequence of events causes an \texttt{ArrayOutOfBoundsException}:
(1) In the main window, click \texttt{Manage content selectors}, which opens a
new dialog; (2) switch to the main window and choose \texttt{Close database}.
Then, (3) switch back to the previously opened dialog and click \texttt{OK}. The
error occurs, because the new opened dialog is started \emph{modeless}, which
allows the user to close the database, although the dialog still suggests the
user to modify the database.

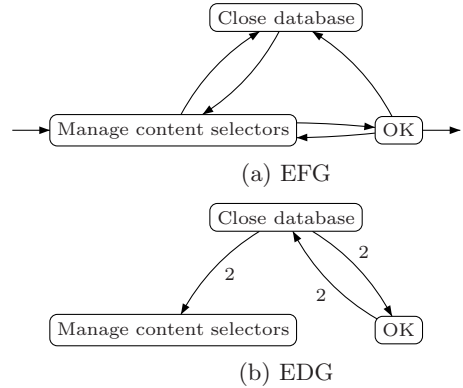
\begin{figure}[h]
\centering
 \subfigure[EFG]{\vbox to 2cm{
  \begin{picture}(60,17)
    \scriptsize
    \gasset{Nadjust=wh,Nadjustdist=1,Nmr=1}
    \node(cd)(30,15){Close database}
    \node[Nmarks=i](mcs)(15,0){Manage content selectors}
    \node[Nmarks=f](ok)(45,0){OK}

    \drawedge[curvedepth=2](cd,mcs){}
    \drawedge[curvedepth=2](mcs,cd){}
    \drawedge[curvedepth=1](mcs,ok){}
    \drawedge[curvedepth=1](ok,mcs){}
    \drawedge[curvedepth=-2](ok,cd){}
  \end{picture}
 }}
 \centering
 \subfigure[EDG]{\vbox to 2cm{
   \begin{picture}(60,17)
    \scriptsize
    \gasset{Nadjust=wh,Nadjustdist=1,Nmr=1}
    \node(cd)(30,15){Close database}
    \node(mcs)(15,0){Manage content selectors}
    \node(ok)(45,0){OK}

    \drawedge[curvedepth=-2](cd,mcs){2}
    \drawedge[curvedepth=2](cd,ok){2}
    \drawedge[curvedepth=2](ok,cd){2}
  \end{picture}
  }}
  \caption{EFG and EDG snippet of JabRef.}
  \label{fig:jabref-efg-edg}
\end{figure}
 
Figure~\ref{fig:jabref-efg-edg} shows the EFG and EDG of JabRef that corresponds
to the found bug. In the event sequence generation for event \texttt{Close
database}, the grey-box approach detects the data dependency to event
\texttt{OK}. This data dependency (weight = 2) consists of a field for JabRef's
metadata, which is written in \texttt{OK} and read in \texttt{Close database}.
Thus, the \abstractTC{} $\langle$\texttt{Close database}, \texttt{OK}$\rangle$
is generated. This \abstractTC{} is converted into an \execTC{}, because there
exists no corresponding path in the EFG. Algorithm~\ref{alg:edg2efg} picks the
shortest path from an initial event to \texttt{Close database}, and the shortest
path between succeeding events to \texttt{OK}, which leads to the following
\execTC{}: $\langle$\texttt{Manage content selectors}, \texttt{Close database},
\texttt{Manage content selectors}, \texttt{OK}$\rangle$. The black-box approach
will be able to detect this failure using a event sequence of length 4. However,
it will first need to generate and execute all possible sequences of length 4.

\textbf{Observation 10:} The third bug was found in Rachota with
Configuration D. The following sequence of events causes a
\texttt{NullPointerException} at restart: (1) Click on \texttt{System settings};
(2) Add a new task (\texttt{Add task}) and leave the text fields blank; (3)
click the OK button (\texttt{OK2}). Then, (4) click on the OK button
(\texttt{OK2}), that writes all tasks to a file. The errors occurs, because the
new added task contains a \texttt{null} value when it is written to the user
settings. Then, a null-reference is returned when the user settings are read,
which is not correctly handled.

\vspace{-0.7cm}

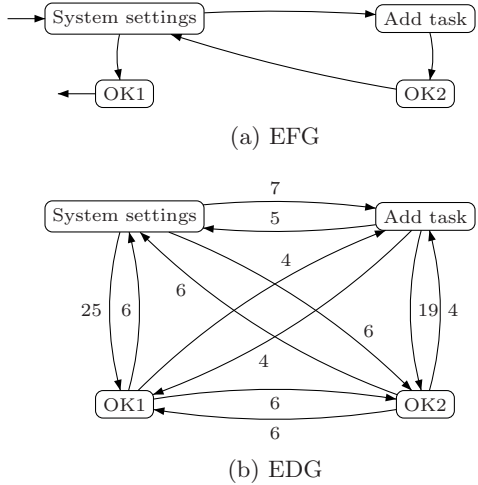
\begin{figure}[h]
\centering

  \subfigure[EFG]{\vbox to 2.3cm{
  \begin{picture}(40,20)
    \scriptsize
    \gasset{Nadjust=wh,Nadjustdist=1,Nmr=1}
    \node[Nmarks=i](ss)(0,10){System settings}
    \node(at)(40,10){Add task}
    \node[Nmarks=f,fangle=180](ok1)(0,0){OK1}
    \node(ok2)(40,0){OK2}

    \drawedge[curvedepth=1](ss,at){}
    \drawedge[curvedepth=1](at,ok2){}
    \drawedge[curvedepth=1](ok2,ss){}
    \drawedge[curvedepth=-1](ss,ok1){}
  \end{picture}  
  }}
  \newline
  \centering 
  \subfigure[EDG]{\vbox to 3.6cm{
  \begin{picture}(40,30)
    \scriptsize
    \gasset{Nadjust=wh,Nadjustdist=1,Nmr=1}
    \node(ss)(0,25){System settings}
    \node(at)(40,25){Add task}
    \node(ok1)(0,0){OK1}
    \node(ok2)(40,0){OK2}

    \drawedge[curvedepth=2](ss,at){7}
    \drawedge[curvedepth=2,ELside=r](at,ss){5}
    \drawedge[curvedepth=-2](at,ok2){19}
    \drawedge[curvedepth=-2,ELside=r](ok2,at){4}
    \drawedge[curvedepth=2](ok2,ok1){6}
    \drawedge[curvedepth=2,ELside=r](ok1,ok2){6}
    \drawedge[curvedepth=-2](ok1,ss){6}    
    \drawedge[curvedepth=-2,ELside=r](ss,ok1){25}

    \drawedge[curvedepth=3,ELpos=75](ss,ok2){6}
    \drawedge[curvedepth=3,ELpos=75](ok2,ss){6}
    \drawedge[curvedepth=3,ELpos=60](at,ok1){4}
    \drawedge[curvedepth=3,ELpos=60](ok1,at){4}   
  \end{picture}
  }}
  \caption{EFG and EDG snippet of Rachota.}
  \label{fig:rachota-efg-edg}
\end{figure}
 
Figure~\ref{fig:rachota-efg-edg} shows the EFG and EDG of Rachota that
corresponds to the bug. When generating \abstractTCs{}, we choose those
sequences with the highest edge values first. For event \texttt{OK2}, the first
\abstractTC{} is $\langle$\texttt{OK2}, \texttt{System settings}$\rangle$, with
a weight of 6. Our second \abstractTC{} is $\langle$\texttt{OK2},
\texttt{OK1}$\rangle$, with a weight of 6. Since this \abstractTC{} is not
allowed to execute in its current form, it is converted into an \execTC{} using
Algorithm~\ref{alg:edg2efg}. Hence, the final \execTC{} that can be run on the
application is $\langle$\texttt{System settings}, \texttt{Add task},
\texttt{OK2}, \texttt{System settings}, \texttt{OK1}$\rangle$. The black-box
approach will be able to detect this failure using an event sequence of length
5. However, the number of event sequences with length 5 would be 6,605,912,
since Rachota consists of 154 events.

\section{Discussion}
\label{sec:discussion}

Regarding the research question of our experiment we found that the answer of
\textbf{Q1} is \textbf{yes}: Considering possible data dependencies between
events lead to fewer event sequences and decreases the time to run all test
cases. The answer of \textbf{Q2} is \textbf{no}: We did not find enough
evidences to show an improvement of the effectiveness.

The main result is that the grey-box approach in most cases produces a lot less
test cases (each generated \abstractTC{} represents one test case) than the
black-box approach. Our initial assumption is that in GUIs several widgets carry
out completely independent tasks. For example, a toolbox usually offers
\emph{save}, \emph{print}, \emph{copy}, \emph{undo/redo} and \emph{find}.
However, \emph{print} is unlikely to have a side effect on all other widgets.
Thus, it is not efficient to test all combinations of \emph{print} plus one
other event (see Configuration B). Here, the grey-box approach can achieve
significant improvements. For Rachota, more \abstractTC{} are generated than in
the black-box approach. The reason is that the events in Rachota have a lot of
dependencies to other events. In this case, more edges in the EDG than in the
EFG are obtained, and thus, more \abstractTC{} are generated. However, we
observe that in the other AUTs the number of EFG edges is higher than the number
of EDG edges.

Increasing the length of event sequences in both approaches (black-box and
grey-box) implies a considerably increase of the generation time. However, if we
compare with the execution time, the generation time itself is not a big issue.
Moreover, in practice, the testing process can be very limited in terms of
resources and time to generate and execute all event sequences and test cases
respectively. In this way, we could adapt our approach to an on-the-fly test
case generation, where a specific timeout is given and parameter like $\tlen$
and $\ttop$ are not fixed, but vary in a range.

Using the grey-box approach, two different bugs were found, that were not found
in the black-box approach, and, there are two main reasons:
(1) all \abstractTCs{} incorporate data dependencies in the application's
bytecode, and (2) the \abstractTCs{} have a non-fixed length. For instance,
while Configuration A, B, and C select events that are directly connected,
Configuration D, E, and F select events based on their data dependencies. Thus,
the \execTC{} length in these grey-box configurations may vary comparing to the
black-box configurations. Further, in the grey-box approach the length of an
\execTC{} can be very long, e.g., if the distance of events in an \abstractTC{},
in terms of intermediate events, is very high in the EFG.

The overall code coverage reported in our experiment is relatively low for
several reasons. For instance, key strokes (\texttt{KeyListener}) and mouse
gestures (\texttt{MouseListener}) are not yet considered, but frequently used in
the application FreeMind, in order to draw a mind-map via mouse interactions.
Support for these events in the GUI ripper and Replayer is scheduled for the
future release of GUITAR. Moreover, the use of random input-data may lead to the
execution default branches in the applications.

\section{Threats to Validity}
\label{sec:threats}

We report 2 threats to internal validity. The first is the experiment
replication. Almost all applications store user settings to the HDD, such as
enabled and disabled toolbars, recently opened files etc. In order to ensure the
precondition (i.e., the system's state) for each run of a test case, it is
important that those user settings have to be deleted before execution.
Otherwise, test cases may mistakenly fail, e.g., a GUI component is not found
due to an existing user setting. In order to decrease this threat to internal
validity we ran the experiment twice and got the same result.

The second is that some applications are strongly connected to the date and time
of their execution. For instance, GUI components like \emph{calendar controls}
are considered in the GUI ripper and in the construction of the EFG. When
replaying the test cases, some of them may fail, because the GUI components are
not recognized anymore (during replaying the calendar control shows a different
date as the calendar control was ripped).

One threat to external validity is the portability of the configurations. For
instance, mobile phones have a different environment and the construction of the
EFG and EDG can be completely different. In principle, there is no reason to
believe that the grey-box approach is not applicable to other platforms. To
generalize the approach to other platforms, we must first port the ripper and
replayer tools. Further, the model implementations have to be adapted to the
corresponding environment. In this way, we believe that our approach can be
generalized to different platforms.

\section{Related Work}
\label{sec:related}

Several approaches for modeling GUI-based applications have been developed for
test case generation.

\textbf{Model-based GUI testing:} Different models can be used for event
sequence generation~\cite{belli2001,nikolai2005,reza2007}. For example, AI
planning techniques are used in~\cite{memon1999}; covering arrays
in~\cite{YuanCohenMemonTSE2011}. Event sequences are generated from these models
and executed as test cases on the GUI to validate its behavior. In the grey-box
approach, the EDG, created by analyzing bytecode, is used to generate event
sequences. In~\cite{ganov2009}, symbolic execution is used to find adequate
inputs for event sequences. While symbolic execution is a powerful technique to
find precise input values, it's applicability is limited due to the complexity
of the used algorithms. In contrast, the grey-box approach only tries to
identify simple data dependencies without tracking the actual value of fields,
and thus it is applicable for reasonably sized applications.
In~\cite{mcmaster2008}, a method to dynamically observe a program's behavior at
execution time is presented. Instead of analyzing the source code, an analysis
of the call stack at run-time is performed. Event Sequences are then generated
such that a minimum set covers a maximum possible set of program execution
paths. In~\cite{DBLP:conf/icse/MarianiPRS11,pezze2012} the AutoBlackTest
approach is presented, which constructs a GUI model by learning how the GUI
interacts with the system functionalities. Then, the tool selects an executable
and non-redundant test suite. They also compare with the GUITAR approach.
However, we could not empirically compare with AutoBlackTest since it is not
available at the moment. In \cite{YuanMemonIST2010}, feedback obtained by
executing an event sequence is used to generate an improved test suite. It is an
iterative method where GUI run-time feedback is used instead of source code
information. In \cite{karam2006} the execution of a GUI-based application is
represented as a sequence of events and output states. A state graph for the GUI
is built which makes it possible to apply code based testing methods to GUIs.

\textbf{Byte, Binary and Source Code Analysis:} Many tools are available for
reachability analysis and state space exploration of programs using the byte,
binary, or source code. For example, JavaPathFinder~\cite{groce2002} works at
the Java bytecode level to identify deadlocks, assertion violations and other
properties of the program using heuristics for reducing the state space
explosion. Soot~\cite{DBLP:conf/cc/LhotakLH04} is designed to be a framework to
allow researchers to experiment with analyses and optimizations of Java
bytecode. In the grey-box approach, we are interested in detecting sequences of
events, which eventually bring the system to a failure state. Hence, we decide
to implement a light-weight bytecode analysis, which can be enhanced by the
support of alternative tools.

\textbf{Search-based testing:} In~\cite{DBLP:conf/icse/BaresiM10} a search-based
testing technique is proposed. Unit tests for Java classes and methods are
generated by looking for tests that satisfy given heuristics. Another approach
using search-based testing is proposed in~\cite{DBLP:conf/sigsoft/FraserA11}.
Heuristics are used to generate test cases that violate automated test oracles.
In the grey-box approach, a data dependency can be seen as a heuristic, which
helps to sample the user-level model (EFG) more efficiently.

The grey-box approach is similar to the generation of sequences of method calls,
e.g., in libraries. However, when system testing an application through its GUI,
not all methods (event handlers) may be available. For instance, a check box is
likely to have no separate event handler, which changes the value from
\emph{selected} and \emph{deselected}, once a user clicks this check box. This
behavior may be implemented in the GUI framework and is not existing in the
application itself. Without a user-level model it is difficult to generate a
proper event sequence, if the value of the check box is evaluated in a further
event handler (method) within the application, while it was changed in the GUI
framework. Further, providing precise input for data-bound widgets, e.g., for an
event handlers that governs a text box, is tough. Transferring input-data to a
text box during test case execution, e.g., via reflection, may violate an
invariant of the class. For instance, when the text box is disabled with regard
to the event-flow, and does not accept any input.

\section{Conclusion and Future Work}
\label{sec:conclusion}

We presented a new automatic grey-box approach for GUI event sequence
generation. An EFG is generated automatically by observing the GUI at run-time
(black-box). In addition, the application's bytecode is analyzed to find
\emph{data dependency} between event handlers (white-box) to generate a model called
event-dependency graph (EDG). \AbstractTCs{} representing data dependencies are
first generated from the EDG. These are then converted into \ExecTCs{} by
looking up the EFG.

The grey-box approach incorporates 2 main steps: (i) model construction (EFG and
EDG), and (ii) event sequence generation. The approach improves event sequence
generation by producing fewer test cases and avoids generating event sequences
where consecutive events share no data dependencies. Empirical evaluation shows
that the grey-box approach decreases the time to generate event sequences and
the time for executing test cases while retaining coverage.

Utilizing a black-box and a white-box model for the generation of event
sequences is promising: We plan to improve the creation of the models and the
generation of event sequences:

\textbf{Model Creation:} We plan to enhance the analysis of event handlers and
the computation of the weight between two events respectively.
Table~\ref{tab:exp-results} shows the potential for increasing the coverage of
the AUT's. We believe that analyzing conditionals, i.e., if-, switch-, and
loop-statements, can lead to the execution of more lines and branches.
In the long run, the grey-box approach is supposed to provide a framework, where
different black-box and white-box techniques can be plugged to generate event
sequences. For instance, one would like to guide a dynamic symbolic
execution~\cite{DBLP:conf/pldi/GodefroidKS05} based on the EFG, or wrap a GUI
application in a set of parameterized unit
tests~\cite{DBLP:conf/issta/FraserZ11}.

% First, we plan to introduce an interactive tool, which
% enhances the EFG during reverse engineering. A tester can specify precise input
% data for widgets (e.g., for a textbox), and to identify widgets nested deep in
% the GUI. This is required for adding nodes and edges which may have been missed
% by the GUI Ripper.

\textbf{Event Sequence Generation:} Typically, applications contain a subset of
events with a relatively high number of dependent events, e.g., the system
settings are read in many other event handlers. The grey-box approach enables us
to identify these events, which we call \emph{hot spots}. Intuitively, \emph{hot
spots} may be fault prone owing to inter-procedural data dependencies.
In a future work, we plan to specifically analyze \emph{hot spots} while
generating event sequences. More precisely, our event sequences generation uses
the parameters $\tlen$ and $\ttop$ while generating a set of executable test
cases. Considering \emph{hot spots} might be useful to limit the event sequence
generation and test case execution to a specific timeout.
The idea is to spend a certain time on the testing of highly dependent events.

Evaluating the fault detection effectiveness is an important aspect of automated
event sequence generation techniques. As a future work, we consider to evaluate
the effectiveness of sequences generated from the EDG. We will start with
fault-seeded versions of the application, and then naturally occurring faults in
fielded applications. In order to have strong evidence about the experiment, we
plan to execute the \emph{best} configuration with different seeds.

\section*{Acknowledgments}
\label{sec:acknowledgments}
The authors would like to thank Simon Pahl who supported us in the
implementation. This work is partially supported by the research projects EVGUI
and ARV funded by the Macau Science and Technology Development Fund, and the US
National Science Foundation under grant CNS-0855055.

\bibliographystyle{abbrv}
\bibliography{issta2012}

\end{document}